\RequirePackage{fix-cm}
\documentclass[twocolumn,epjc3]{svjour3}  
\smartqed  
\RequirePackage{graphicx}
\journalname{Eur. Phys. J. C}
\usepackage{graphicx}
\usepackage{dcolumn}
\usepackage{bm}

\usepackage{amssymb}
\usepackage{amsmath}
\usepackage{amsfonts}
\usepackage{slashed}
\usepackage{graphicx}
\usepackage{relsize}
\usepackage{color}
\usepackage{extarrows}
\usepackage{cancel}
\usepackage{placeins}
\usepackage{float}
\usepackage{subcaption}
\usepackage{enumitem}
\usepackage{mathtools}

\definecolor{gray}{rgb}{0.5, 0.5, 0.5}

\RequirePackage{chapterbib}

\begin{document}

\title{Geometric Proca with Matter in Metric-Palatini Gravity}


\author{Durmu{\c s} Demir$ \thanksref{addr1,e1}$ 
        \and
        Beyhan Puli{\c c}e$ \thanksref{addr1,e2}$ 
}

\thankstext{e1}{e-mail: durmus.demir@sabanciuniv.edu}
\thankstext{e2}{e-mail: beyhan.pulice@sabanciuniv.edu}

\institute{Faculty of Engineering and Natural Sciences, Sabanc{\i} University, 34956 Tuzla, {\.Istanbul}, Turkey \label{addr1}
}

\date{Received: date / Accepted: date}

\maketitle

\begin{abstract}
In the present work, we study linear, torsion-free metric-Palatini gravity, extended by the quadratics of the antisymmetric part of the Ricci tensor and extended also by the presence of  the affine connection in the matter sector.  We show that this extended metric-Palatini gravity reduces dynamically to the general relativity plus a geometrical massive vector field corresponding to non-metricity of the connection. We also show that this geometric Proca field couples to fermions universally. We derive static, spherically symmetric field equations of this Einstein-geometric Proca theory.  We study possibility of black hole solutions by taking into account the presence of a dust distribution that couples to the geometric Proca. Our analytical and numerical analyses show that the presence of this dust worsens the possibility of horizon formation. We briefly discuss possible roles of this universally-coupled geometric Proca in the astrophysical and collider processes.
\end{abstract}

\section{Introduction}
Metric-Palatini gravity \cite{Palatini} is an extension of the  general relativity (GR) in which the connection and  metric tensor are independent quantities. One motivation for Palatini gravity is that it gives Einstein field equations dynamically with no need to exterior curvature \cite{york,gh}. Another motivation is that it envelops geometrical scalars and vectors \cite{Demir2012}. An independent recent motivation comes from the relevance of Palatini gravity for emergent gravity theories \cite{Demir2019, Demir2021}. Extension of the Palatini gravity with fundamental scalars like the Higgs field leads to natural inflation \cite{bauer-demir1,bauer-demir2}. Extension of the Palatini gravity with higher-curvature terms \cite{Vitagliano2011,Vitagliano2013,Demir2020}, on the other hand, leads to rich astrophysical and cosmological phenomena \cite{Palatini-f(R)}. 

Extension of the Palatini gravity with the metrical curvature sets a new direction. Such extensions, dubbed as ``metric-Palatini gravity" theories \cite{harko2012,Capozziello2013a,Capozziello2015}, have led to novel predictions for dark matter \cite{Capozziello2012a}, wormholes \cite{Capozziello2012b,Rosa2018}, cosmology \cite{Capozziello2012c,Rosa2017}, static Universe \cite{Bohmer2013}, flat rotation curves \cite{Capozziello2013c}, vacuum solutions \cite{Danila2018}, weak field phenomenology \cite{Rosa2021,Bombacigno2019}, stability of black holes \cite{Rosa2020} and thermodynamics \cite{Azizi2015}. 

The framework in the present work is set by the metric-Palatini gravity theories \cite{harko2012,Capozziello2013a,Capozziello2015} that are 
\begin{eqnarray}
\label{our-model}
&& {\it (i)}\ \text{linear in the affine scalar curvature},\nonumber\\
&& {\it (ii)}\ \text{torsion-free, and}  \\
&& {\it (iii)}\ \text{quadratic in the antisymmetric part}\nonumber \\
&&\quad\quad\,\text{of the affine curvature}\ \cite{Vitagliano2011,Vitagliano2013,Demir2020}. \nonumber
\end{eqnarray}
We call this framework the {\it extended metric-Palatini gravity} (EMPG).  The EMPG involves the affine connection $\Gamma^\lambda_{\mu\nu}$ (independent of the metric) and the usual Levi-Civita connection ${}^g\Gamma^\lambda_{\mu\nu}$ (deriving from the metric). The difference between $\Gamma^\lambda_{\mu\nu}$ and ${}^g\Gamma^\lambda_{\mu\nu}$ is a rank (1,2) tensor \cite{Demir2012}, and this tensor field defines the geometrodynamics beyond the GR within a given action. The EMGP possesses two geometric tensor structures beyond the curvature. The first is the  nonmetricity tensor $Q_{\lambda\mu\nu}\equiv -{}^\Gamma \nabla_\alpha g_{\mu\nu}$, where ${}^\Gamma \nabla_\alpha$ is the covariant derivative with respect to the affine connection  $\Gamma^\lambda_{\mu\nu}$. The second is the  torsion tensor $T^\lambda_{\mu\nu}\equiv \Gamma_{\mu\nu}-\Gamma_{\nu\mu}$. Torsion will be taken zero throughout the paper. In fact, it has been shown that torsion-free Palatini gravity quadratic in the antisymmetric part of the  Ricci tensor reduces to the GR plus a massive vector field theory \cite{Demir2020}. The massive vector in question is the non-metricity vector $Q_\mu \equiv \frac{1}{4} Q _{\mu \nu}^{~~~\nu}$. The GR plus this massive vector field forms an Einstein-Proca system \cite{Buchdahl1979, Tucker1996, Dereli1996,Obukhov1997,Vitagliano2010}, where coupling of the Proca field to matter is what shall be emphasized in the present work. 

The Einstein-Proca system has been analyzed for  Reissner-Nordstr\"{o}m (RN) type spherically-symmetric vacuum solutions  \cite{Tresguerres1995a, Tresguerres1995b, Tucker1995, Obukhov1996, Vlachynsky1996, Macias1999} (not considering the geometrical Proca studied in the present work). The particular role played by the Proca field  has been revealed \cite{Bekenstein1971, Bekenstein1972, Adler1978}, and this reveleation has been extended to spherically-symmetric static solutions \cite{Frolov1978, Gottlieb1984, Leaute1985}, with further analysis of the horizon  \cite{Ayon1999, Obukhov1999, Toussaint2000}. In the present work, for definiteness, we will call the dynamical system at hand ``Einstein -- geometric Proca" (EGP) system to emphasize the purely geometrical origin of the massive vector field  \cite{Demir2020, Buchdahl1979, Tucker1996, Dereli1996,Obukhov1997,Vitagliano2010}. 

In the present work, we will elucidate the geometrical Proca field and perform a detailed study of its interactions with  matter fields in the framework of the EMPG model. These interactions arise if the  affine connection appears also in the matter sector
\cite{bauer-demir1, bauer-demir2, affine-VE,affine-CC,affine-RE}. And to this end one notes that fermion kinetic terms  involve the affine connection through the spin connection, and this involvement makes all fermions couple to the non-metricity tensor universally \cite{Demir2020,Buchdahl1979, Tucker1996, Dereli1996}. 

In Sec. II we give a detailed analysis of the EMGP, starting with a ghost-free Lagrangian more general defined in (\ref{our-model}) above. In this framework, we show how EMGP dynamically reduces to involve only the non-metricity vector (starting with a rank-3 non-metricity tensor) as a massive Proca field. In other words, we show how EMGP reduces to EGP dynamically.  Our metric-Palatini gravity setup in Sec. II will have the general structure to have nontrivial implications various gravitational phenomena like black holes, neutron stars, gravitational waves, and particle scatterings at colliders \cite{Demir2020}. 

In Sec. III,  we give a static and spherically-symmetric solution of the EGP model. We perform a detailed analysis of the EGP setup by employing a perfect fluid ansatz for the fermions. (Fermions are the only fields which the Proca field couples directly). In our analysis, we consider different coupling regimes.    

In Sec. IV, we apply the spherically-symmetric static solution of Sec. III to black holes.  We show that there do not exist true black hole solutions due mainly to the absence of the event horizon (even though their asymptotic behaviour suggests the  RN black hole behavior). We specialize to spherically-symmetric dust distribution by considering both the neutral (without any coupling to the geometric Proca) and charged (with couplings to the geometric Proca) matter (mainly the fermions).   To this end, we analyze the spherically-symmetric static solutions of the EGP system for the dusty black holes, {\it i.e.} the black hole surrounded by a geometrically charged dust distribution. In this case, the geometric Proca field couples to the geometrical charge density of the dust  distribution. One of the motivations to analyze the EGP model for a dusty black hole solution is the recent observation of the European Southern Observatory's Very Large Telescope Interferometer \cite{Rosas2021}. By observing the center of the galaxy Messier 77, it is detected that a thick ring of dust is hiding a supermassive black hole, and determination of the dust by these electromagnetic observations can be useful for determining the geometrical charge distribution.  The dusty black hole solutions reveal that the presence of matter fields and their couplings to the geometric vector fields have significant affects in that solutions with geometrically-charged dust distributions are shifted drastically away from the dusty solutions with geometrically-neutral dust distributions.

In Sec. V we conclude.

\section{Extended Metric-Palatini Gravity with Matter}
In accordance with the conditions stated in (\ref{our-model}),  the EMPG framework is defined by the following action (with the metric signature $(-,+,+,+)$)
\begin{align}
S\left[g, \Gamma, \Psi \right] &= \int d^4 x\, \sqrt{-g} \Bigg\{\frac{M^2}{2} g^{\mu \nu}{R}_{\mu\nu} ({^g}\Gamma) \nonumber \\ 
&+ \frac{\overline{M}^2}{2} g^{\mu\nu}{\mathbb{R}}_{\mu\nu}(\Gamma) -  \frac{\xi}{4}  {\overline{\mathbb{R}}}_{\mu\nu}(\Gamma) {\overline{\mathbb{R}}}^{\mu\nu}(\Gamma) \nonumber \\ 
&+ {\mathcal{L}}\left(g, \Gamma, \Psi\right)\Bigg\} 
\label{action-mag}
\end{align}
in which 
\begin{align}
\label{ricci1}
{R}_{\mu\nu}({}^g\Gamma)  =  \partial_{\lambda}{}^g\Gamma^{\lambda}_{\mu\nu} - 
\partial_{\nu}{}^g\Gamma^{\lambda}_{\lambda\mu} + {}^g\Gamma^{\rho}_{\rho\lambda} {}^g\Gamma^{\lambda}_{\mu\nu}  - {}^g\Gamma^{\rho}_{\nu\lambda} {}^g\Gamma^{\lambda}_{\rho\mu},
\end{align}
is the metrical Ricci curvature of the  Levi-Civita connection ${}^g\Gamma^\lambda_{\mu\nu}$. Obviously,
the affine Ricci tensor $\mathbb{R}_{\mu\nu}(\Gamma)$, following from the Riemann tensor as $\mathbb{R}_{\mu\nu}(\Gamma)\equiv  {\mathbb{R}}^{\lambda}_{\mu\lambda\nu}(\Gamma)$), is obtained by replacing  the Levi-Civita connection ${}^g\Gamma$  with the affine connection $\Gamma$ in (\ref{ricci1}). The matter fields, collectively denoted as $\Psi$, are described by the Lagrangian  ${\mathcal{L}}\left(g, \Gamma, \Psi\right)$, with possible presence of the affine connection $\Gamma$. The second affine Ricci tensor $\overline{\mathbb{R}}_{\mu\nu}(\Gamma)$, obtained from the Riemann tensor by the contraction ${{\overline{\mathbb{R}}}}_{\mu\nu}(\Gamma) \equiv {\mathbb{R}}^{\lambda}_{\lambda\mu \nu}(\Gamma)$, possessed the simple expression 
\begin{align}
\label{ricci2}
{{\overline{\mathbb{R}}}}_{\mu\nu}(\Gamma) =  \partial_{\mu}\Gamma^{\lambda}_{\lambda\nu} - \partial_{\nu}\Gamma^{\lambda}_{\lambda\mu},
\end{align}
as the antisymmetric part of the affine Ricci tensor ${{\mathbb{R}}}_{\mu\nu}(\Gamma)$. It does obviously vanishes identically in the GR limit ($\Gamma \rightarrow {^g}\Gamma$).

The action (\ref{action-mag}) falls in the class of  metric-Palatini gravity theories \cite{harko2012,Capozziello2013a,Capozziello2015}. Its first term (proportional to $M^2$) corresponds to the Einstein-Hilbert term in the GR. Its second term (proportional to ${\overline{M}}^2$) means that we consider the linear case of the metric-Palatini gravity [$g^{\mu \nu}{\mathbb{R}}_{\mu\nu} (\Gamma)$ in place of $f(g^{\mu \nu}{\mathbb{R}}_{\mu\nu} (\Gamma))$]. Its third term (proportional to $\xi$) gives the extension of the metric-Palatini gravity with the antisymmetric part of the affine Ricci curvature \cite{Vitagliano2011,Vitagliano2013,Demir2020}. Finally, its fourth term emphasizes the presence of the affine connection in the matter sector (mainly the fermion kinetic terms) \cite{Demir2020, Buchdahl1979, Tucker1996, Dereli1996}.

The gravitational actions like (\ref{action-mag}) can always be extended by higher-power curvature invariants thanks to general covariance.  One can consider therefore ghost-free functional forms like $f(g^{\mu \nu}{R}_{\mu\nu} ({^g}\Gamma))$ \cite{fR-metric} or \\ $f(g^{\mu \nu}{\mathbb{R}}_{\mu\nu} (\Gamma))$ \cite{fR-affine}, where the latter has already been analyzed in the metric-Palatini gravity \cite{harko2012,Capozziello2013a,Capozziello2015}. It is possible to consider also Ricci-squared (as well as Riemann-squared) terms, which have been studied in both the metrical \cite{Stelle77,Carroll2004,Clifton2006,Middleton2010,Nojiri2010,Nojiri2017} and Palatini formalisms \cite{Vitagliano2010,Borowiec96,Li2007,Olmo2009,Olmo2011,Allemandi2005}. These terms are known to  contain gravitational ghosts. However, it is known that projective symmetry prevents ghost-like instabilities in Ricci-based gravity theories \cite{Jimenez2019}.  The EMPG action (\ref{action-mag}) would also contain ghosts if it were containing quadratics of ${\mathbb{R}}_{\mu\nu}(\Gamma)$ and ${\mathbb{R}}^{\alpha}_{\mu\beta\nu}(\Gamma)$. The reason is that such quadratics would lead to  metrical curvature terms like ${\mathbb{R}}_{\mu\nu}({}^g\Gamma)$ or ${\mathbb{R}}^{\alpha}_{\mu\beta\nu}({}^g\Gamma)$, which are known to carry ghosts \cite{Stelle78}. The cross terms ${\mathbb{R}}_{\mu\nu}({}^g\Gamma){\mathbb{R}}^{\mu\nu}(\Gamma)$ would be another possibility but it is known that it is not possible to prevent ghosts in that case, too \cite{Koivisto2013}. Nevertheless, one here notes that there are exceptions to this, as exemplified by the metric-affine theory with higher-spin fields \cite{sezgin} as well as the higher-curvature gravity theories with propagating torsion \cite{sezgin1979, sezgin1981}. There are also studies on stability against radiative corrections in metric-affine gravity \cite{Marzo2021-1,Marzo2021-2}. It is also known that the torsion-free Ricci based theories with a vector field degree of freedom are also ghost-free \cite{Jimenez2019}. In view of all these ghosty structures, the EMPG action in (\ref{action-mag}) stands out as a nominal ghost-free setup. It can certainly be generalized by functional forms like $f(g^{\mu \nu}{R}_{\mu\nu} ({^g}\Gamma))$ \cite{fR-metric} or $f(g^{\mu \nu}{\mathbb{R}}_{\mu\nu} (\Gamma))$ \cite{fR-affine}. These ghost-free higher-curvature actions are known to lead to the GR plus a geometrical scalar, and are expected to enrich the EMPG by adding geometrical scalars to already-present geometrical Proca. In the present work, we will limit ourselves to the action (\ref{action-mag}) as we are interested primarily in the couplings of the geometric Proca with matter fields (not those of the geometrical scalars). 

It proves convenient to analyze the EMPG action (\ref{action-mag}) by decomposing the affine connection as follows \cite{Demir2020}
\begin{align}
\Gamma^{\lambda}_{\mu\nu} = {}^g \Gamma^{\lambda}_{\mu\nu} +  \Delta^\lambda_{\mu\nu}
\label{decomp}
\end{align}
where $\Delta^\lambda_{\mu\nu}$ is a symmetric rank-(1,2) tensor field. The two Ricci curvature tensors accordingly take the  forms
\begin{align}
\label{ricci-nonmetricity}
{\mathbb{R}}_{\mu\nu}(\Gamma)  &= {\mathbb{R}}_{\mu\nu}({}^g \Gamma) + {}^g \nabla_{\lambda} \Delta^{\lambda}_{\mu\nu} - {}^g \nabla_{\nu} \Delta^{\lambda}_{\lambda\mu} \nonumber \\ 
&+ \Delta^{\rho}_{\rho\lambda} \Delta^{\lambda}_{\mu\nu}  - \Delta^{\rho}_{\nu\lambda} \Delta^{\lambda}_{\rho\mu}~, \\
{{\overline{\mathbb{R}}}}_{\mu\nu} (\Gamma) &=  \nabla_{\mu} \Delta^\lambda_{\lambda \nu} - \nabla_{\nu} \Delta^\lambda_{\lambda \mu}
\end{align}    
under which the action (\ref{action-mag}) takes the form 
\begin{align}
&S\left[g, \Delta, \Psi \right] = \int d^4 x\, \sqrt{-g} \Bigg \{ \frac{M^2 + \overline{M}^2}{2} g^{\mu\nu}{\mathbb{R}}_{\mu\nu}({^g}\Gamma) \nonumber \\
&- \frac{\xi}{4} g^{\mu \alpha} g^{\nu \beta} (\nabla_{\mu} \Delta^\lambda_{\lambda \nu} - \nabla_{\nu} \Delta^\lambda_{\lambda \mu}) (\nabla_{\alpha} \Delta^\lambda_{\lambda \beta} - \nabla_{\beta} \Delta^\lambda_{\lambda \alpha}) \nonumber \\
&+\frac{\overline{M}^2}{2} g^{\mu \nu} (\Delta^\rho_{\rho \lambda} \Delta^\lambda_{\mu \nu} - \Delta^\rho_{\nu \lambda} \Delta^\lambda_{\rho \mu}) + \mathcal{L} (g, {}^g \Gamma, \Delta, \Psi) \Bigg \}
\label{action-decomp}
\end{align}
in which $\nabla_\alpha$ is covariant derivative with respect to the Levi-Civita connection so that $\nabla_\alpha g_{\mu\nu} = 0$.  The tensor field $\Delta^\lambda_{\mu\nu}$  is seen to appear both in the geometrical and matter sectors. It proves useful to define
\begin{eqnarray}
M^2 + \overline{M}^2 \equiv \frac{1}{\kappa}
\end{eqnarray}
where $\kappa=8\pi G_N$, $G_N$ being the Newton's constant. It is possible to bring the EMPG action (\ref{action-decomp}) into a familiar form  by expressing $\Delta^\lambda_{\mu\nu}$ in terms of the non-metricity tensor as follows \cite{Buchdahl1979, Tucker1996, Dereli1996,Obukhov1997,Vitagliano2010}  (torsion is zero everywhere)
\begin{align}
\Delta^{\lambda}_{\mu\nu} = \frac{1}{2} g^{\lambda \rho} ( Q_{\mu \nu \rho } + Q_{\nu \mu \rho } - Q_{\rho \mu \nu} )
\label{decomp-nonmetricity}
\end{align}
in which
\begin{align}
Q_{\lambda \mu \nu} = - {}^\Gamma \nabla_{\lambda} g_{\mu \nu}
\end{align}
is the non-metricity tensor. It proves useful to define also the non-metricity vector
\begin{align}
Q_\mu = \frac{1}{4} Q _{\mu \nu}^{~~~\nu}
\label{nm-vector}
\end{align}
which will prove useful in the  dynamical equations in the sequel.

The affine connection $\Gamma^\lambda_{\mu\nu}$ or the geometrical tensor $\Delta^\lambda_{\mu\nu}$ takes part in the matter action 
$\mathcal{L} (g, {}^g \Gamma, \Delta, \Psi)$ through the fermions kinetic terms. This is due to the spin connection in the  covariant derivative of the spinor fields in curved spacetime \cite{Kosmann66, Hurley94, Fatibene96,Adak2002}
\begin{align}
\label{spin-connection}
{}^\Gamma \nabla_\mu \psi = (\partial_\mu + \frac{1}{4} \omega_\mu^{a b} \gamma_a \gamma_b) \psi
\end{align}
in which the flat spacetime Clifford algebra gives
\begin{align}
\gamma_a \gamma_b = \eta_{ab} + 2 \sigma_{ab}     
\end{align}
with the Lorentz generator
\begin{align}
\sigma_{ab} = \frac{1}{4} [\gamma_a, \gamma_b].
\end{align}
Using the decomposition of the affine connection in  (\ref{decomp-nonmetricity}), the affine covariant derivative in (\ref{spin-connection}) decomposes as
\begin{align}
{}^\Gamma \nabla_\mu \psi = (\nabla_\mu + \frac{1}{2} Q_\mu) \psi    
\label{cov-deriv-spinor}
\end{align}
to explicitly contain the non-metricity vector $Q_\mu$ defined in (\ref{nm-vector}). This decomposition proves that each and every fermion field couples to the non-metricity vector universally. 
 
Expressing  $\Delta^\lambda_{\mu\nu}$ in terms of the non-metricity tensor $Q_{\lambda\mu\nu}$ via the relation 
(\ref{decomp-nonmetricity}), the EMPG action  (\ref{action-decomp}) turns to the action of  the non-metricity tensor. In fact, this new action remains stationary against variations in the affine connection $\Gamma^\lambda_{\mu \nu}$ provided that
\begin{align}
&{}^\Gamma \nabla_\lambda (\sqrt{-g}g^{\mu \nu}) - {}^\Gamma \nabla_\sigma (\sqrt{-g}g^{\sigma (\nu}) \delta^{\mu)}_\lambda \nonumber \\
&+ \frac{\xi}{\overline{M}^2} \Big [ {}^\Gamma \nabla_\sigma (\sqrt{-g} {\overline{\mathbb{R}}}^{\sigma \mu}) \delta^\nu_\lambda + {}^\Gamma \nabla_\sigma (\sqrt{-g} {\overline{\mathbb{R}}}^{\sigma \nu}) \delta^\mu_\lambda  \Big ]\nonumber\\ &- \frac{1}{\overline{M}^2} \sqrt{-g} \frac{\delta \mathcal{L}}{\delta \Gamma^\lambda_{\mu \nu}} = 0
\label{eom-connection-1}
\end{align}
which governs the dynamics of  $Q_{\lambda\mu\nu}$. Taking trace of 
(\ref{eom-connection-1}), replacing divergences in it, and using the spinor covariant derivative (\ref{cov-deriv-spinor}) in computing the variation $\frac{\delta \mathcal{L}}{\delta \Gamma^\lambda_{\mu \nu}}$ of the matter Lagrangian, the non-metricity equation of motion (\ref{eom-connection-1}) takes the form 
\begin{align}
&{}^\Gamma \nabla_\lambda (\sqrt{-g} g^{\mu \nu}) - \frac{2 \xi}{3 \overline{M}^2} \Big ( {}^\Gamma \nabla_\sigma (\sqrt{-g} \overline{\mathbb{R}}^{\mu \sigma}) \delta^\nu_\lambda \nonumber \\
&+ {}^\Gamma \nabla_\sigma (\sqrt{-g} \overline{\mathbb{R}}^{\nu \sigma}) \delta^\mu_\lambda \Big ) \nonumber \\
&+ \frac{1}{6 \overline{M}^2} \sqrt{-g} \Big ( \overline{\psi} \gamma^\mu \psi \delta^\nu_\lambda + \overline{\psi} \gamma^\nu \psi \delta^\mu_\lambda \Big ) = 0
\label{eom-connection-2}
\end{align}
such that its contraction 
\begin{align}
g_{\mu \nu} {}^\Gamma\nabla_\lambda (\sqrt{-g} g^{\mu \nu}) &=  \frac{4 \xi}{3 \overline{M}^2} g_{\lambda \nu} {}^\Gamma\nabla_\mu (\sqrt{-g} \overline{\mathbb{R}}^{\nu \mu}) \nonumber \\
&-  \frac{1}{3 \overline{M}^2} \sqrt{-g} g_{\nu \lambda} \overline{\psi} \gamma^\nu \psi
\end{align}
leads to the equation
\begin{align}
\nabla_\mu \overline{\mathbb{R}}^{\mu \nu} - \frac{3 \overline{M}^2}{\xi} Q^\nu = - \frac{1}{4 \xi} \overline{\psi} \gamma^\nu \psi
\label{eom-Q}
\end{align}
via the relation 
\begin{align}
{}^\Gamma \nabla_\mu (\sqrt{-g} \overline{\mathbb{R}}^{\nu \mu} ) = \sqrt{-g} ~ \nabla_\mu \overline{\mathbb{R}}^{\nu \mu}.
\end{align}
As a result, the relation (\ref{eom-Q}) for the non-metricity vector implies that the contracted equation of motion (\ref{eom-connection-2}) enjoys this specific relation
\begin{align}
Q_{\lambda \mu  \nu} = 2  Q_\lambda g_{\mu \nu} - 2  Q_\mu g_{\nu \lambda} - 2  Q_\nu g_{\mu \lambda}
\label{tensor2vector}
\end{align}
expressing the non-metricity tensor in terms of the non-metricity vector. This means that the EMPG  dynamics involves the curved metric $g_{\mu\nu}$ and the non-metricity vector $Q_{\mu}$ as the two dynamical fields, and the relation (\ref{tensor2vector}) takes the definition of $\Delta _{\mu \nu}^{\lambda}$ in (\ref{decomp-nonmetricity}) to the new form
\begin{align}
\Delta _{\mu \nu}^{\lambda} = - 3 Q^\lambda g_{\mu \nu} +  Q_\nu \delta_\mu^\lambda +  Q_\mu \delta_\nu^\lambda
\end{align}
under which the EMPG  action (\ref{action-decomp}) reduces to the following action of the GR plus a massive vector field theory with matter (as in our previous work \cite{Demir2020})
\begin{align}
S[g,Y,\psi] &= \int d^4 x \sqrt{-g} \Bigg \{ \frac{1}{2 \kappa} R(g) - \frac{3 \overline{M}^2 }{4 \xi} Y_{\mu} Y^{\mu} \nonumber \\
&- \frac{1}{4} Y_{\mu \nu} Y^{\mu \nu} + \frac{1} {4 \sqrt{\xi}} \overline{\psi} \gamma^\mu \psi Y_\mu  \nonumber \\
&+ {\mathcal{L}}_{rest} (g,{}^g \Gamma,\psi) \Bigg \}
\label{action-EGP}
\end{align}
which is nothing but the aforementioned EGP model. Here,  $R(g)\equiv g^{\mu\nu}{\mathbb{R}}_{\mu\nu}({}^g\Gamma)$ is the metrical curvature scalar, 
$Y_\mu \equiv 2 \sqrt{\xi} Q_{\mu}$
is the normalized canonical vector field generated by the affine connection, $Y_{\mu \nu} = \partial_\mu Y_\nu -\partial_\nu Y_\mu $ is the field strength tensor of $Y_{\mu}$, and ${\mathcal{L}}_{rest}\left(g, {}^g\Gamma, \Psi \right)$ is part of the matter Lagrangian that does not involve $Y_{\mu}$. (As noted before, generalizations like $f(g^{\mu \nu}{R}_{\mu\nu} ({^g}\Gamma))$ \cite{fR-metric} or $f(g^{\mu \nu}{\mathbb{R}}_{\mu\nu} (\Gamma))$ \cite{fR-affine} would produce a geometrical scalar in (\ref{action-EGP}) setup above.)  In the action (\ref{action-EGP}), the  $Y_\mu$ mass is set by the mass scale $\overline{M}$ while the coupling of  $Y_\mu$ to the fermions are set only by the parameter $\xi$. This  structure of parameter space allows to set the fundamental scale of the gravity correctly while the vector field mass and its interaction strength vary independently in a wide range ($\xi: 0 \leftrightarrow 1$ and ${\overline{M}}^2: 0 \leftrightarrow 1/\kappa$). (This parameter space, much wider than in \cite{Demir2020}, results from the metric-Palatini structure of the EMPG action \cite{harko2012,Capozziello2013a,Capozziello2015}.) It is clear from the fourth term of the action (\ref{action-EGP}) that $Y_{\mu}$  couples universally only to fermions (not to the Higgs and gauge bosons).  This interaction of $Y_{\mu}$ with fermions will certainly have important implications for various physical phenomena. 
It is clear that the motion equation (\ref{eom-connection-2}) for the non-metricity vector leads to the following motion equation for  $Y_\mu$
\begin{align}
\nabla_\mu Y^{\mu \nu} - M^2_Y Y^\nu = - g_Y J^\nu   
\label{eom-Y}
\end{align}
where $J^\nu = \overline{\psi} \gamma^\nu \psi$ is the fermion current,
\begin{eqnarray}
M_Y^2 = \frac{3 {\overline{M}}^2}{2\xi}
\end{eqnarray}
is the $Y_\mu$ mass-squared, and
\begin{eqnarray}
g_Y = \frac{1}{4 \sqrt{\xi}}  
\end{eqnarray}
is the $Y_\mu$ coupling constant, resembling the gauge coupling in gauge theories though $Y_\mu$ is by origin not a gauge field at all.

\section{Spherically-Symmetric Static EGP System, with Perfect Fluid}

To set the stage, we bring the EGP action (\ref{action-EGP}) into the compact form
\begin{align}
S[g,Y] &= \int d^4 x \sqrt{-g} \Bigg \{ \frac{1}{2 \kappa} R(g) -  \frac{1}{4} Y_{\mu \nu} Y^{\mu \nu} - \frac{1 }{2} M_Y^2 Y_{\mu} Y^{\mu}  \nonumber \\
&+ g_Y Y_\mu J^\mu + \mathcal{L}_{rest} \Bigg \}
\label{action-spherical}
\end{align}
which we will analyze below in search for spherically-symmetric static solutions. Its  variation with respect to the metric $g_{\mu\nu}$ leads to the Einstein field equations
\begin{align}
R_{\mu \nu} ({}^g \Gamma) - \frac{1}{2} R(g) g_{\mu \nu} = \kappa (T_{\mu \nu}^Y + T_{\mu \nu}^{rest})
\label{Einstein-eqns}
\end{align}
wherein the two energy-momentum tensors at the right-hand side are given by 
\begin{align}
&T_{\mu \nu}^Y = Y_{\alpha \mu } Y_{\beta \nu} g^{\alpha \beta}-\frac{1}{4} Y_{\alpha \beta} Y^{\alpha \beta} g_{\mu \nu} + M_Y^2 ( Y_\mu Y_\nu \nonumber \\
&- \frac{1}{2} Y_\alpha Y^\alpha g_{\mu \nu}) + 2 g_Y ( Y_\mu J_\nu - \frac{1}{2} Y_\alpha J^\alpha g_{\mu \nu} ), \\
&T_{\mu \nu}^{rest} = \mathcal{L}_{rest} g_{\mu \nu} - 2\frac{\delta \mathcal{L}_{rest}}{\delta g_{\mu \nu}}.
\label{en-moms}
\end{align}
In an attempt to find spherically-symmetric static solutions of the field equations (\ref{Einstein-eqns}) in space coordinates $(r,\theta, \phi)$, we put forth the ansatz
\begin{align}
g_{\mu \nu} = \text{diag}(-f^2(r),\frac{g^2(r)}{f^2(r)} ,r^2, r^2 \sin ^2 \theta)  
\label{metric-sss}
\end{align}
in which $g^2(r)$ measures the deviation from the Scwarzschild-like form.

Having done with the metric, the geometric Proca field $Y_\mu$  obeying the equation of motion (\ref{eom-Y}) can be taken  as a
purely time-like field 
\begin{align}
Y_\mu = \frac{u(r)}{r} \delta_\mu^0    
\label{proca-sss}
\end{align}
in agreement with  spherically-symmetric background. Here $u(r)$ measures deviation from the Coulombic limit. With this time-like vector, gravitational and  geometric-Proca parts of the EGP model are described by three real functions $f(r), g(r)$ and $u(r)$. What remains is spefication of their sources. The source of the metric tensor $g_{\mu\nu}$, taken to be a perfect fluid, has the form (see the definitions (\ref{en-moms}) above)
\begin{align}
T_{\mu \nu}^{rest} = (\rho_M + p_M) v_\mu v_\nu + p_M g_{\mu \nu}
\label{em-sss}
\end{align}
where $\rho_M$ and $p_M$ are the energy density and  pressure of the perfect fluid, respectively and the four-velocity of the perfect fluid satisfies  $v_\mu v^\mu = -1$. In this context, source of the geometric Proca $Y_\mu$ takes the form 
\begin{align}
J_\mu = \rho_{C} v_\mu
\label{current-sss}
\end{align}
where $\rho_{C}$ is the geometrical charge density of the matter distribution. This current is a mean-field approximation to the   fermion current defined below (\ref{eom-Y}), and sources the  geometric Proca field $Y_\mu$. 



Now, for the ansatze in (\ref{metric-sss}), (\ref{proca-sss}), (\ref{em-sss}) and (\ref{current-sss})  the total energy-momentum tensor takes the form
\begin{align}
T_{\mu \nu} &= \frac{1}{2 r^2 g^2 f^2} \Big [-f^2(u^\prime-\frac{u}{r})^2 + M_Y^2 g^2 u^2\Big ] g_{11} \delta_\mu^1 \delta_\nu^1 \nonumber \\ 
&- \frac{1}{2 r^2 g^2 f^2} \Big [f^2(u^\prime-\frac{u}{r})^2 + M_Y^2 g^2 u^2\Big ] (-g_{00} \delta_\mu^0 \delta_\nu^0 \nonumber \\
&+ g_{22} \delta_\mu^2 \delta_\nu^2 +  g_{33} \delta_\mu^3 \delta_\nu^3)  \nonumber \\ 
&+ \Big [\rho_M + g_Y \frac{\rho_{C}}{f} \frac{u}{r}  \Big ] (-g_{00} \delta_\mu^0 \delta_\nu^0) \nonumber \\ 
&+ \Big [ p_M + g_Y \frac{\rho_{C}}{f} \frac{u}{r}  \Big ] (g_{11} \delta_\mu^1 \delta_\nu^1 + g_{22} \delta_\mu^2 \delta_\nu^2 + g_{33} \delta_\mu^3 \delta_\nu^3)
\end{align}
with which the Einstein field equations (\ref{Einstein-eqns}) can be solved component-by-component. 
The $(\mu \nu = 11)$ and $(\mu \nu = 00)$  components lead, respectively, to the ordinary differential equations 
\begin{align}
\label{E11}
2 r (f^2)^\prime + 2(f^2-g^2) &= \kappa \Big [ \frac{M_Y^2 g^2}{f^2}  u^2-(u^\prime - \frac{u}{r})^2 \nonumber \\
&+ 2 p_M r^2  g^2 +  \frac{2 g_Y \rho_{C} r g^2  }{f} u \Big ] , \\
r (g^2)^\prime &=  \kappa \Big [ \frac{M_Y^2 g^4}{f^4} u^2 + \frac{r^2 g^4}{f^2} (\rho_M + p_M) \nonumber \\
&+ \frac{2g_Y \rho_{C} g^4}{f^3} u  \Big ]
\label{E11E00}
\end{align}
where prime ($^\prime$) stands for derivatives with respect to $r$. 

By a similar analysis, the equation of motion (\ref{eom-Y}) of the geometric Proca turns to  
\begin{align}
\label{eom-Y-spherical}
u^{\prime \prime} =  \frac{M_Y^2 g^2}{f^2} u  + \frac{g^\prime}{g} \Big ( u^\prime - \frac{u}{r} \Big ) + g_Y \frac{\rho_{C} r}{f}.
\end{align}
under the ansatze (\ref{metric-sss}), (\ref{proca-sss}),  and (\ref{current-sss}). 

Needless to say, the system of equations (\ref{E11}), (\ref{E11E00}) and (\ref{eom-Y-spherical}) is a coupled nonlinear ordinary differential equations set. It is hard to solve analytically, and we will therefore resort to numerical techniques. To this aim, we carry the equations into gravitational units to obtain dimensionless equations system via the following dimensionless quantities:
\begin{align}
&\hat{r} := \kappa^{-1/2} r,~ 
{\hat M}_Y^2 := \kappa M_Y^2,~
\hat{p}_M := \kappa^2 p_M,~
\hat{\rho}_M := \kappa^2 \rho_M, \nonumber \\
&\hat{\rho}_{C} := \kappa^{3/2} \rho_{C}
\end{align}
so that the original differential equations  (\ref{E11}), (\ref{E11E00}) and (\ref{eom-Y-spherical}) take the following form:
\begin{align}
\label{E11-dimensionless}
2 \Big ( \hat{r} \frac{d f^2}{d \hat{r}} + f^2 - g^2 \Big) &= {\hat M}_Y^2 \frac{g^2}{f^2} u^2 - \Big ( \frac{d u}{d \hat{r}} - \frac{u}{\hat{r}}\Big )^2 \nonumber \\
&+ 2 \hat{p}_M \hat{r}^2  g^2 + \frac{2 g_Y  \hat{\rho}_{C} \hat{r} g^2}{f} u , \\
\label{E11E00-dimensionless}
\hat{r} \frac{d g^2}{d \hat{r}} &= {\hat M}_Y^2 \frac{g^4}{f^4} u^2 + \frac{\hat{r}^2 g^4}{f^2} (\hat{\rho}_M + \hat{p}_M) \nonumber \\
&+ \frac{2 g_Y \hat{\rho}_{C} \hat{r} g^4}{f^3} u, \\
\label{eom-Y-dimensionless}
\frac{d^2 u}{d \hat{r}^2} &= {\hat M}_Y^2 \frac{g^2}{f^2} u + \frac{1}{2 g^2} \frac{d g^2}{d \hat{r}} \Big ( \frac{d u}{d \hat{r}} - \frac{u}{\hat{r}} \Big ) \nonumber \\
&+ g_Y \frac{\hat{\rho}_{C} \hat{r}}{f}.
\end{align}
Here, the dimensionless radial coordinate $\hat{r}$ measures the distance from the origin in units of the Planck length $\kappa^{1/2}$. This is a system of coupled nonlinear ordinary differential equations describing the dynamics of metric and geometrical Proca field in the presence of a matter distribution with both energy and geometrical charge densities. The coupling of $Y_\mu$ to the geometrical charge density of the matter arises in the EGP model naturally, and we will show in the next section that it actually has inevitable effects on the spherically-symmetric static solutions of the EGP system. This matter distribution something not discussed before, it is a new topic brought about by the present work. 

The EGP model, even as a spherically-symmetric static system described by (\ref{E11-dimensionless}), (\ref{E11E00-dimensionless}) and (\ref{eom-Y-dimensionless}), can have important implications in numerous astrophysical systems like black holes, neutron stars, gamma ray bursters, magnetars and as such. The spherically-symmetric static black hole solutions of  the RN type of have been studied in the literature \cite{Obukhov1999, Toussaint2000}. The thing is that in these papers (and in other relevant work) matter distribution (with or without geometrical charge) has not been taken into account. 
Here, in the present work, in Sec.\ref{sec-black-hole-soltns} below, we will focus on black hole solutions with a geometrical dust distribution around. We call the solutions "dusty black hole solutions". We will show that the coupling of $Y_\mu$ to the geometrical charge density of the dust distribution has significant effects on the behaviour of the solutions.

\section{Dusty Black Hole Solutions \label{sec-black-hole-soltns}}

In this section we analyze the dimensionless system of equations (\ref{E11-dimensionless}), (\ref{E11E00-dimensionless}) and (\ref{eom-Y-dimensionless}) in the presence of a dust distribution. We take for the dust
\begin{align}
\hat{p}_M &= 0, \\
\hat{\rho}_M &= \frac{\hat{M}_D}{\hat{r}^3}, \\
\hat{\rho}_{C} &= \frac{Q_D}{\hat{r}^3}
\label{dust-sss}
\end{align}
in which $\hat{M}_D = \sqrt{\kappa} M_D$ and $Q_D$ are the dimensionless mass and the geometrical charge, respectively. (We chose this distribution as a nominal structure. It is possible to consider more general distributions depending on the physical system under concern.) 

Before starting the numerical analysis, it proves useful to determine the asymptotic behaviors of the fields.   The approximate solutions will be a combination of the solutions small  $\hat{r} \rightarrow 0$ and large $\hat{r} \rightarrow \infty$ distance solutions. Firstly, in the asymptotic Minkowski spacetime for which $f^2=g^2=1$  at $\hat{r} \rightarrow \infty$, the geometric Proca assumes the solution
\begin{align}
u &= Q_B~ e^{- {\hat M}_Y \hat{r}} - \frac{g_Y Q_D}{2} \Big ( e^{{\hat M}_Y \hat{r}} {\rm Ei}(-{\hat M}_Y \hat{r}) \nonumber \\
&+ e^{-{\hat M}_Y \hat{r}} {\rm Ei}({\hat M}_Y \hat{r}) \Big ) 
\end{align}
in which the first term is the homogeneous solution corresponding to the Yukawa potential emanating from a black hole of geometrical charge $Q_B$. The second term, the particular solution, gives the Proca field as sourced by a geometrically-charged (charge $Q_D$) dust distribution defined in (\ref{dust-sss}). It vanishes when $Q_D = 0$, as expected. Here, 
\begin{eqnarray}
{\rm Ei}(x) = - \int_{-x}^\infty e^{-t} d(\log t)
\end{eqnarray}
is the exponential integral function. (This particular solution will be different for different charge distributions $\rho_C$.)

Around the origin,  $\hat{r} \rightarrow 0$, effects of the geometric Proca mass is negligibe ($e^{- {\hat M}_Y \hat{r}} \simeq 1- {\hat M}_Y \hat{r}$). Then, in the absence of the mass density  $\hat{\rho}_M$ and the charge density $\hat{\rho}_{C}$ of the dust distribution,  the black hole mass ${\hat M}_B$ would be the only gravitating source. This leads to the solution of $g^2 = c_0^2 $ by (\ref{E11E00-dimensionless}) where $c_0$ is a constant. Then, under the same conditions, the geometric Proca becomes $u\simeq Q_B$ by (\ref{eom-Y-dimensionless}). In consequence, including the energy and geometric charge densities, the metric function acquires the solution  
\begin{align}
\label{fsquare-approx}
f^2 = c_0^2 \Big ( 1 - \frac{2 {\hat M}_B}{\hat{r}} + \frac{Q_B^2}{2 c_0^2 \hat{r}^2} -  g_Y \frac{  Q_B Q_D }{ \hat{r}^2} \Big ) 
\end{align}
which holds near the origin with a Schwarzschild part proportional to ${\hat M}_B$ and the RN-type contributions proportional to $Q_B^2$ and $g_Y$. This reveals the effect of charge distribution of the dust on the spacetime structure. 

In light of  the discussions above, the asymptotic solutions ($\hat{r} \rightarrow \infty$)  read as follows: 
\begin{align} 
\label{fsquare_inf}
f^2_\infty &\rightarrow  c_0^2 \Big ( 1 - \frac{2 {\hat M}_B}{\hat{r}_\infty} + \frac{Q_\infty^2}{2 c_0^2 \hat{r}_\infty^2} - g_Y  \frac{  Q_\infty Q_D }{\hat{r}_\infty^2} \Big ), \\
\label{gsquare_inf}
g^2_\infty  &\rightarrow c_0^2, \\
\label{u_inf}
u_\infty &\rightarrow Q_\infty e^{- {\hat M}_Y  \hat{r}_\infty} - \frac{ g_Y Q_D}{2} \Big ( e^{{\hat M}_Y \hat{r}_\infty} {\rm Ei}(-{\hat M}_Y  \hat{r}_\infty) \nonumber \\
&+ e^{-{\hat M}_Y  \hat{r}_\infty} {\rm Ei}({\hat M}_Y  \hat{r}_\infty) \Big ), \\
\label{uprime_inf}
u^\prime_\infty &\rightarrow - Q_\infty {\hat M}_Y  e^{- {\hat M}_Y  \hat{r}_\infty} - g_Y Q_D \Big [ \frac{1}{\hat{r}_\infty} \nonumber \\
&+ \frac{1}{2} {\hat M}_Y  \Big ( e^{{\hat M}_Y  \hat{r}_\infty} {\rm Ei}(-{\hat M}_Y  \hat{r}_\infty) \nonumber \\
&- e^{-{\hat M}_Y  \hat{r}_\infty} {\rm Ei}({\hat M}_Y  \hat{r}_\infty) \Big ) \Big ]
\end{align}
where $Q_\infty$ is the asymptotic geometrical charge of the black hole. The asymptotic metric function $f^2$ (\ref{fsquare_inf}) is in the form of a RN type solution with an additional term which emerges due to the coupling of $Y_\mu$ to the geometrical charge density of the dust distribution. The Schwarzchild solution is naturally expected as an approximate solution far away from the gravitating mass $M_B$; however, we also keep the $\hat{r}^{-2}$ terms in this asymptotic solution of $f^2$ (\ref{fsquare_inf}). 

Having set all three dynamical equations in gravitational units, we now turn to their numerical integrations. To this end, we take
\begin{eqnarray}
{\hat M}_Y=1,~c_0 = 1,~Q_B=Q_\infty = 1,~\hat{r} = 30
\end{eqnarray}
in the asymptotic functions (\ref{fsquare_inf}), (\ref{gsquare_inf}), (\ref{u_inf}) and (\ref{uprime_inf}). Below, we consider geometrically-neutral ($Q_D=0$) and geometrically-charged ($Q_D\neq 0$) dusts separately to determine effects of matter on spherically-symmetric static solutions.

\subsection{Geometrically-Neutral Dust}

\begin{figure}
\begin{minipage}{.5\linewidth}
\centering
\subfloat[]{\label{main:a}\includegraphics[scale=.45]{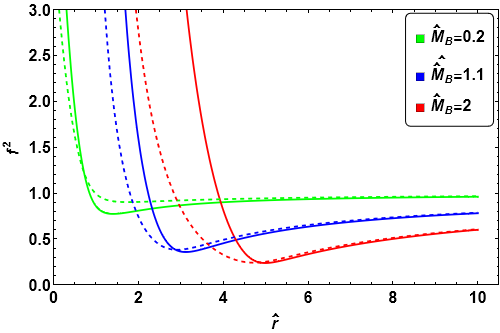}}
\end{minipage} \\
\begin{minipage}{.5\linewidth}
\centering
\subfloat[]{\label{main:b}\includegraphics[scale=.45]{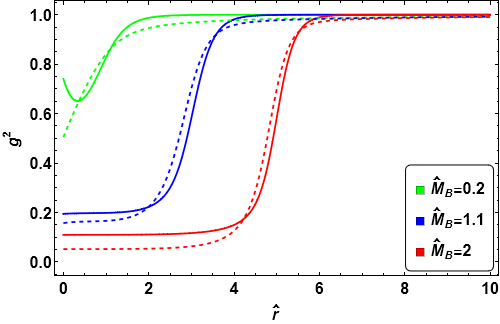}}
\end{minipage}\par\medskip
\centering
\subfloat[]{\label{main:c}\includegraphics[scale=.45]{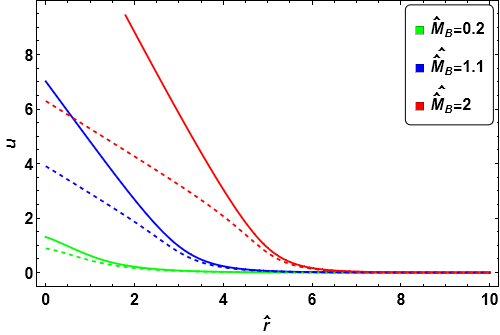}}
\caption{The spherically-symmetric static EGP system as described by the metric functions 
$f^2$ and $g^2$, and by the geometric-Proca function $u$. The solid (dashed) curves correspond to dustless (dusty) case. The dust has zero geometrical charge, that is, $Q_D=0$. The black hole mass is varied over $\hat{M}_B=0.2, 1.1, 2$, and it is observed that effects of the geometrically-neutral dust is pronounced at large $\hat{M}_B$ and small $\hat{r}$ regimes.}
\label{figures_matter}
\end{figure}

In this subsection, we perform a numerical analysis of the  effects of the geometrically-neutral ($Q_D=0$) dust.  Depicted in FIG.\ref{figures_matter} are $f^2$ (panel (a)), $g^2$ (panel (b)), and $u$ (panel (c)) as a function of ${\hat{r}}$ without dust (full curves) and with dust (dashed curves) for different values of the black hole mass ${\hat M}_B = 0.2$ (green curve), ${\hat M}_B = 1.1$ (blue curve), and ${\hat M}_B = 2.0$ (red curve). The dust has zero geometrical charge ($Q_D=0$). It is clear from the figure that the metric functions $f^2({\hat r})$, $g^2({\hat r})$ and $u({\hat r})$ remain nonzero and non-negative for the entire parameter ranges. Comparison of the full curves with the dashed curves reveals that effects of geometrically-neutral dust get pronounced at low ${\hat r}$ and high ${\hat M}_B$  values. These effects do, however, never bring these functions to zero, which means that black hole solutions in the EGP system develop a horizon neither with dust nor without dust. This means that there are no true black hole solutions. 

\subsection{Geometrically-Charged Dust}
\begin{figure}
\begin{minipage}{.5\linewidth}
\centering
\subfloat[]{\label{main:a}\includegraphics[scale=.45]{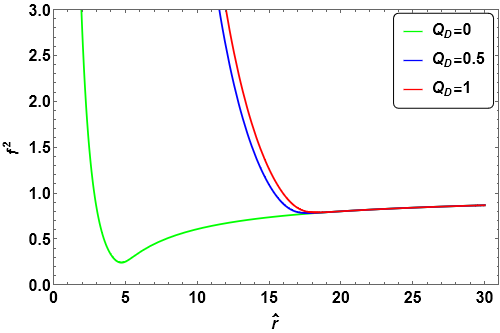}}
\end{minipage} \\
\begin{minipage}{.5\linewidth}
\centering
\subfloat[]{\label{main:b}\includegraphics[scale=.45]{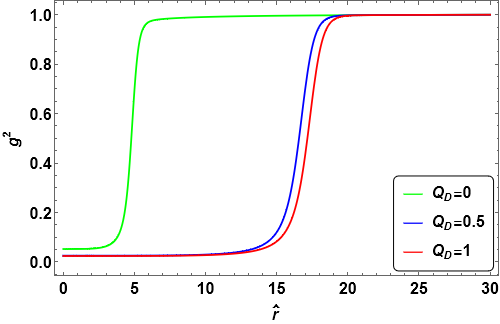}}
\end{minipage}\par\medskip
\centering
\subfloat[]{\label{main:c}\includegraphics[scale=.45]{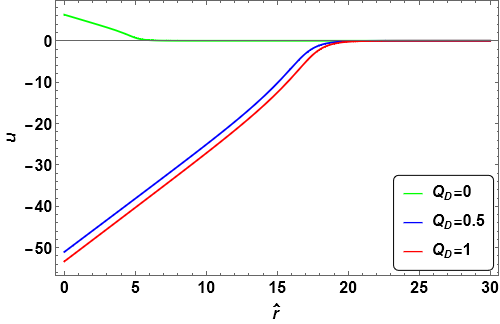}}
\caption{The spherically-symmetric static EGP system as described by the metric functions $f^2$ and $g^2$, and by the geometric-Proca function $u$. The plots are produced by setting $\hat{M}_B=2$ and $\hat{M}_D = 0.1$, and $g_Y=0.1$. The green, blue and red curves correspond to $Q_D=0$, $Q_D=0.5$ and $Q_D=1$, respectively. It is clear from the figure that the presence of the charged dust modifies $f^2$ and $g^2$ and $u$ drastically from the homogeneous one (green) to particular (blue, red) solutions. This is due to dominance of $\rho_C$ at small ${\hat r}$.}
\label{figures_gY}
\end{figure}

In this subsection, we perform a numerical analysis of the  effects of the geometrically-charged ($Q_D\neq 0$) dust. In this particular case, the geometric Proca field has a source set by $\rho_C$. The solution is composed of the uniform solution in Subsection A plus the particular solution set by $\rho_C$. Depicted in FIG.\ref{figures_gY} are $f^2$ (panel (a)), $g^2$ (panel (b)), and $u$ (panel (c)) as a function of ${\hat{r}}$ for geometrically-neutral dust ($Q_D=0$ with green curve) and geometrically-charged dust ($Q_D\neq0$ with blue and red curves). What is spectacular about these metric and geometric Proca configurations is that the particular solution gets abruptly shifted from the homogeneous solution by the presence of a geometrically-charged dust distribution ($Q_D\neq 0$). The reason for this is mainly the choice of $\rho_C$ in that $\rho_C \propto 1/{\hat r}^3$ causes the field configurations to take much larger values at low ${\hat r}$ but merges with the homogeneous solution at large ${\hat r}$. The reason for clustering of the solutions is that shifts in $Q_D$ does not change the overall  ${\hat r}$ dependence. All these properties are confirmed by the approximate solution of $f^2$ in (\ref{fsquare-approx}) in that geometric charge effects (involving $Q_B^2$ and $Q_B Q_D$) grow as $1/{\hat r}^2$ at small ${\hat r}$ and surpass the Schwarzschild solution. 
In fact, for small charges like $Q_D \sim 10^{-9}$ the particular solutions (blue, red) are seen to get closer to the homogeneous (green) solution. These ${\hat r}$  dependencies reveal that neither the dustless nor the dusty EGP system can give rise to black holes despite the fact that the approximate solutions exhibit RN behavior. Indeed, as seen from FIG.\ref{figures_gY} panel $(a)$, charged dust causes push $f^2$ away from the zero-axis and diminishes therefore possibility of developing a horizon.  

\section{Discussions and Conclusion}

In this paper, we have performed a systematic study of the EMPG model in the linear ghost-free limit in which quadratic and higher-order curvature terms are all dropped. It falls in the class of linear, torsion-free, metric-Palatini gravity theories 
\cite{harko2012,Capozziello2013a,Capozziello2015}, with the extensions that  a term  quadratic in the antisymmetric part of the affine curvature \cite{Vitagliano2011,Vitagliano2013,Demir2020} exists
and the matter action involves the affine connection explicitly \cite{Buchdahl1979, Tucker1996, Dereli1996,Obukhov1997,Vitagliano2010}. 

As we have shown in Sec. II, metric-Palatini gravity theory reduces to the GR plus a geometrical massive vector theory, which we call the Einstein-Geometric Proca (EGP) theory. The EGP model differs from similar models in the literature by its explicit involvement of the direct coupling between the geometric Proca field and the fermions in the theory. To emphasize, it turns out that quarks and leptons (not the Higgs and gauge bosons) couple to the geometrical vector $Y_\mu$ directly, universally and in an Abelian gauge field fashion.

The EGP system provides a novel geometrodynamical framework. It can set the stage for diversely different physical phenomena. Here, we would like to comment on its few salient effects for completeness of the discussions.
\begin{enumerate}
    \item {\bf EGP  in astrophysical media}. These systems involve neutron stars, black holes, magnetars and as such. Implications of the EGP system for such media are determined by solving the EGP field equations (\ref{eom-Y}) and (\ref{Einstein-eqns}). Each system requires a specific structure in terms of time dependence, isotropy and anisotropy. The key point is that the fermions (electrons, protons, neutrons) making up the astrophysical object (a neutron star, for instance) have not only the usual electromagnetic, weak and strong interactions but also the geometric Proca interaction. This additional interaction can cause cooling of stars or instability of neutron stars or novel couplings of different fluidic components in a given astrophysical object.
    
    Our analyses in Sec. III and IV are an illustrative example of what implications the EGP theory can have for spherically-symmetric static geometries. To that end,  we constructed in Sec. III static spherically-symmetric field equations by representing matter fields  by a perfect fluid, and discussed the (im)possibility of black hole type solutions in the presence of dust with or without geometrical charge. Our numerical integration of the EGP field equations showed that possibility of developing a horizon gets lesser and lesser in the presence of a charged dust. 

\item  {\bf EGP at particle colliders}. The geometric Proca field $Y_\mu$, which couple universally to all the leptons and quarks, can give cause to novel signals or event rates or event distributions at collider experiments if its mass lies near the collider reach ($M_Y \gtrsim {\rm TeV}$). In fact, it falls in the general classification of ``$Z^\prime$ models" \cite{Zprime-dad-ilk,Zprime-dad-kane}. It can have non-trivial effects at high-energy colliders \cite{Zprime-dad-nkp,Zprime-LHC,Zprime-LHC-2}.
It can play a role also in the recent fifth force data  \cite{Pulice2021}. Its implications were briefly mentioned in \cite{Demir2020} as part of the dark matter detection problem. Practically speaking, all $Z^\prime$ effects or events like the Drell-Yann production and similar fermion scattering events, can be reanalyzed in view of the universal $Y_\mu$ couplings. Deep down, it will set an interesting example of a geometric field taking part in collider processes like the LHC experiments or future ILC experiments. 

\end{enumerate}

These two are the foremost effects of the EGP system, with matter. The EGP system can affect also the early Universe as a Planckian-mass heavy field that seeds certain anistropies. Our study in this work should make it clear that the EGP can lead to novel physical effects in astrophysical, cosmological and collider settings.

\section*{Acknowledgements}
The work of B.P. is supported by T{\"U}B{\.I}TAK B{\.I}DEB-2218 national postdoctoral fellowship  programme. The authors are grateful to conscientious reviewer for their useful criticisms and suggestions.

\end{document}